\documentclass[pra,twocolumn,showpacs,preprintnumbers,superscriptaddress,amsmath,amssymb]{revtex4}

\usepackage{graphicx}
\usepackage{dcolumn}
\usepackage{bm}
\usepackage{color}

\newcommand{\vc}{\vec}

\newcommand{\LDW}{\ensuremath{L_{\text{DW}}}}
\newcommand{\EDW}{\ensuremath{E_{\text{DW}}}}
\newcommand{\EB}{\ensuremath{E_{\text{B}}}}
\newcommand{\Beff}{\ensuremath{\vc{B}^{\text{eff}}}}
\newcommand{\Bext}{\ensuremath{\vc{B}^{\text{ext}}}}
\newcommand{\mat}{\boldsymbol}
\newcommand{\low}{\ensuremath{\Delta_{\text{low}}}}
\newcommand{\high}{\ensuremath{\Delta_{\text{high}}}}

\date{\today}

\begin{document}

\title{Thermally activated magnetization reversal in monoatomic
  magnetic chains on surfaces studied by classical atomistic
  spin-dynamics simulations}

\author{David S. G. Bauer} \affiliation{Institut f\"ur Festk\"orperforschung, Institute for
  Advanced Simulation, and JARA, Forschungszentrum J\"ulich, D-52425
  J\"ulich, Germany}
\author{Phivos Mavropoulos}\email{Ph.Mavropoulos@fz-juelich.de}
\affiliation{Institut f\"ur Festk\"orperforschung, Institute for
  Advanced Simulation, and JARA, Forschungszentrum J\"ulich, D-52425
  J\"ulich, Germany} 
\author{Samir Lounis} \affiliation{Department of Physics and Astronomy,
University of California Irvine, 
California 92697, USA}
\author{Stefan Bl\"ugel} \affiliation{Institut f\"ur Festk\"orperforschung, Institute for
  Advanced Simulation, and JARA, Forschungszentrum J\"ulich, D-52425
  J\"ulich, Germany}

\begin{abstract}
  We analyze the spontaneous magnetization reversal of supported
  monoatomic chains of finite length due to thermal fluctuations via
  atomistic spin-dynamics simulations. Our approach is based on the
  integration of the Landau-Lifshitz equation of motion of a
  classical spin Hamiltonian at the presence of stochastic forces. The
  associated magnetization lifetime is found to obey an Arrhenius law
  with an activation barrier equal to the domain wall energy in the
  chain. For chains longer than one domain-wall width, the reversal is
  initiated by nucleation of a reversed magnetization domain primarily
  at the chain edge followed by a subsequent propagation of the domain
  wall to the other edge in a random-walk fashion. This results in a
  linear dependence of the lifetime on the chain length, if the
  magnetization correlation length is not exceeded. We studied chains
  of uniaxial and tri-axial anisotropy and found that a tri-axial
  anisotropy leads to a reduction of the magnetization lifetime due to
  a higher reversal attempt rate, even though the activation barrier
  is not changed.
\end{abstract}

\maketitle

\section{Introduction}

Advances in experimental techniques in recent years have made possible
the controlled growth and characterization of magnetic chains on
non-magnetic crystal surfaces. Such chains can be made as thin as
monoatomic, i.e., their cross-section consists of a single atom, while
they are usually a few nanometers or tens of nanometers long; they
grow, e.g., at terrace step edges, in the ``trenches'' of (110)
surfaces or as inclusions in surface alloys
\cite{Gambardella02,Hammer03,Yan05,Vindigni06,Serrate10,Wei09,Honolka09}, so that their
structure is rather stable, and they can be studied by a number of
spin-sensitive techniques including spin-polarized scanning tunneling
microscopy or x-ray magnetic circular dichroism. Magnetic chains, as
all magnetic nanostructures, bear technological relevance due to the
prospect of miniaturization of magnetic bits for information
storage. Particularly appealing, however, in chains is that their
geometry is defined by only one parameter, the length, which makes it
easier to interpret their physical properties \cite{footnote}.

At such small sizes magnetic states at low-temperature equilibrium
consist of a single domain and the possibility of information storage
relies solely on the bistability of the magnetization, which is a
consequence of magnetocrystalline anisotropy. At low enough
temperature the direction of magnetization is trapped close to a local
energy minimum for some time before it is reversed by thermal
flutuations to another minimum, pointing along the opposite
direction. Above some characteristic \emph{blocking} temperature
$T_{\text{B}}$, however, there is a transition to a situation where
the magnetization fluctuations are too intense for a stable state to
be formed. As the size of the nanomagnets is below the thermodynamic
limit, the transition is not a genuine phase transition but rather a
smooth crossover.

These effects are strongly size dependent. In many cases, larger
system size leads to higher blocking temperature, reaching the Curie
temperature in the thermodynamic limit. However, in two-dimensional
\cite{Mermin66} and even more in one-dimensional systems, upon
increasing the system size $L$, the entropy leads the magnetic state
to a multi-domain structure at even low temperatures. The
characteristic length, at which this effect sets in, is the
magnetization correlation length $\xi(T)$; thus, the effect of
magnetic bistability in chains is physically limited by the
requirement $L< \xi(T_{\text{B}})$.

The temperature-driven magnetization reversal dynamics of
nano-particles has been tradidionally studied within the
Stoner-Wohlfarth model \cite{Stoner48,Brown63}, where the particle
magnetization is approximated by a single ``macro-spin'' of rigid
magnitude. However, it is known that excited states departing from the
macro-spin approximation can play a significant role even at sizes of
a few nanometers \cite{Rohart10},
while another mechanism sets in, namely reversal through domain nucleation and 
domain wall propagation \cite{Gambardella04};
the latter mechanism has been seen also in  elongated particles \cite{Bode04}, and should be even
more relevant in chains \cite{Braun06}.

Spin dynamics simulations of magnetic nanostructures are based on a
parametrization of the spin-dependent part of the total energy and
usually follow one of two paths: either that of a quantum Heisenberg
model or the one of a classical Heisenberg model. It is well-known
that the quantum model exhausts computational resources at relatively
small system sizes, as the Hilbert space increases exponentially with
the number of atoms, therefore for system sizes of 50 or 100 atoms
only the classical model is practically available. However, there are
also physical situations where the quantum Heisenberg model is not
applicable, in particular magnetic systems on metallic surfaces. The
quantum model is based on the assumption of half-integer or integer
spins that interact with each other. This is the case with spins on an
insulating surface, such as Mn chains on CuN \cite{Hirjibehedin06}. On
the other hand, magnetic atoms on metallic surfaces are characterized
by non-integer or non-half-integer spin values, which are actually
average values of the spin operator on superposition states of the
magnetic atoms with the substrate continuous spectrum. In this
picture, where substrate electrons hop on and off the magnetic atoms,
the spin states are blurred out compared to the sharp eigenstates of
the quantum Heisenberg model. In addition, damping of the magnetic
excitations into electron-hole pairs emerges, with concequences in
g-shift and frequency-dependent magnetization lifetime
\cite{Balashov09,Lederer67,Muniz03,Lounis10,Khajetoorians10}.  This,
together with temperature effects, causes decoherence of the magnetic
states and thus the quantum Heisenberg model is not applicable any
more; possibly a classical model, founded on the principles of
adiabatic spin dynamics \cite{Antropov96}, is better suited in this
case. Actually the classical atomistic spin dynamics is based on the
Landau-Lifshitz equations of motion, with the point-dependent
magnetization being replaced by atom-dependent magnetic moments, and
is a widely used tool for the study of atom-dependent nanomagnetism
\cite{Antropov97,Skubic08}.

Having this in mind, we use a classical spin-dynamics simulation
method, implemented in our {\tt juSpinx} code, to study the time
evolution of the magnetization in monoatomic chains.  Assuming a
single-ion anisotropy tensor $\mat{K}_i$, inter-atomic exchange
coupling constants $J_{ij}$ between atomic spin moments $\vc{M}_i$ and
$\vc{M}_j$, and an external magnetic field $\Bext$, the spin-dependent
excitation energy of the system is parametrized by a classical
Heisenberg Hamiltonian of the form
\begin{equation}
H = -\frac{1}{2}\sum_{ij}J_{ij}\vc{M}_i\cdot\vc{M}_j -
\sum_{i}\vc{M}_i\cdot\mat{K}_i\vc{M}_i - \sum_{i}\Bext
\cdot \vc{M}_i,
\label{eq:Heisenberg}
\end{equation}
where it is implied that $J_{ii}=0$, while the factor $\frac{1}{2}$
compensates for the double-counting in the summation; $i$ and $j$
denote the atoms in the chain. We set henceforth $|\vc{M}_i|=1$, so
that $\vc{M}_i$ are to be understood as unit vectors along the atomic
moment directions, with the modulus of the moments absorbed in the
parameters $J_{ij}$, $\mat{K}_i$ and $\Bext$. Dipole-dipole
interactions are neglected here, as we are interested in systems of
not more than 100 atoms in size. In order to ensure bistability, the
single-ion anisotropy should favour a particular axis; the simplest
such case is that $\mat{K}_i$ is the same for each atom and its
eigenvalues are $(K_x,K_y,K_z)$, with $K_x=K_y=0$ and $K_z=K>0$,
i.e. the anisotropy is uniaxial, so that second term of the
Hamiltonian takes the form $-\sum_i K M_{i;z}^2$, with the $z$-azis
appropriately chosen along the low-energy direction. In the limit
$K/|J_{ij}|\gg 1$, there can be a further simplification, as the
system behaviour is well-described by the Ising model and it follows
Glauber dynamics \cite{Glauber63}. This condition is met e.g. in
single chain magnets \cite{Bernot06} consisting usually of transition
metal atoms that are stabilized and linked in a chain form by
molecular ligands.  However, for metallic magnetic chains on surfaces
the magnetic anisotropy can take large values that are of the order of
a few meV, while the exchange coupling is typiclaly one or two orders of
magnitude stronger \cite{Mokrousov07}.  In this case, an appropriate classical
approximation to the spin dynamics is the stochastic Landau-Lifshitz
equation of motion:
\begin{equation}
\hbar\frac{\partial\vc{M}_i}{\partial t} = -\vc{M}_i\times \Beff_i -\lambda
\vc{M}_i \times (\vc{M}_i\times\Beff) - \vc{M}_i\times \vc{f}_i(t)
\label{eq:LL}
\end{equation}
where we have defined the effective magnetic field acting on the spin
at the atomic site $i$ as
\begin{eqnarray}
\Beff_i &=& -\frac{\partial H}{\partial \vec{M}_i } \\
&=&
\Bext + \sum_j J_{ij}\vc{M}_j  \nonumber\\
&+&  2 \left[
K_{i,x}(\vc{M}_i \cdot \hat{x})\, \hat{x} 
+K_{i,y}(\vc{M}_i \cdot \hat{y})\, \hat{y} 
+K_{i,z}(\vc{M}_i \cdot \hat{z})\, \hat{z}
\right], \nonumber
\end{eqnarray}
and $\lambda$ is a damping parameter leading the system to equilibrium
at zero temperature. The temperature $T$ enters via a stochastic
fluctuating force $\vc{f}_i(t)$, as described in Appendix A. In
Eq.~(\ref{eq:LL}), the first term on the right-hand side is derived
directly from the Hamiltonian (\ref{eq:Heisenberg}), while the damping
and temperature terms are phenomenologically added assuming
interaction of the spins with a temperature bath.

Our focus in this work is on the temperature-induced magnetization
reversal of finite-length chains. It is implied that the length of the
chain, $L$, is considerably smaller than the correlation length
$\xi(T)$, otherwise the magnetic configuration consists of many
domains, and the bistability loses its meaning. We seek to identify
the lifetime $\tau$ of a magnetic state before it is flipped to the
opposite magnetization direction. The flip is caused by thermal
effects only and we are interested in a ``loss of memory'', therefore
we set $\Bext=0$. It is expected, and found in
the simulation, that $\tau(T)$ follows to a good approximation an
Arrhenius law,
\begin{equation}
\tau(T) = \tau_0 \exp(\EB/k_{\rm B}T),
\label{eq:Arrhenius}
\end{equation}
where $\EB$ is interpreted as an activation energy barrier and
$1/\tau_0$ as an attempt frequency; $k_{\rm B}$ is the Boltzmann
constant 
(details on the method of calculation of $\tau$ are given in
Appendix A). 
We find that $\tau_0$ 
shows a deviation from a constant only at high temperatures, 
where the magnetization does not spend any appreciable time at a state 
before being reversed.
What is important, however, is the dependence of $\EB$ and
$\tau_0$ on the properties of the chain. We keep a minimal model,
considering only nearest-neighbour exchange coupling $J=J_{ij}$ and in
most cases the same anisotropy $K$ at each site (except when
considering edge-anisotropy effects), whence the energy- and
time-scales of the system are defined by $J$, $K$, and
$\lambda$. Then, a domain wall that is formed in the chain will have a
width of approximately $\LDW=2\sqrt{J/K}$ and a formation energy of
$\EDW=2\sqrt{2JK}$ (estimated at the continuum limit \cite{Aharoni00}
of Eq.~\ref{eq:Heisenberg}). As we explain below, our main findings
are categorized in the following regimes: (i) $L\ll\LDW$: the chain
then behaves like a monodomain particle, as laid out already in the
theory of Stoner, Wohlfarth and Brown \cite{Stoner48,Brown63}, with
the reversal being controlled by a barrier of $\EB=LK$. (ii)
$L\gg\LDW$: The magnetization reversal occurs due to a domain
nucleation and subsequent domain wall propagation; the barrier
entering Eq.~(\ref{eq:Arrhenius}) is then $\EB=\EDW$. Regime (ii)
becomes relevant also for shorter chains as the anisotropy increases,
especially above the so-called Ising limit $K/J>1$, when the domain
wall becomes atomically sharp; then even very short chains of a few
atoms in length behave as in regime (ii). These results are presented
in Section \ref{sec:uni}. Furthermore, in Section \ref{sec:multi} we
investigate the possibility of different anisotropy parameters in all
three axes (i.e., we include also an in-plane anisotropy); in this
case, entropic arguments imply that the attempt rate $1/\tau_0$ can be
strongly affected, and this is verified by our simulations. In Section
\ref{sec:real} we discuss our findings for realistic system parameters
of Co chains on Pt. We conclude with a summary in Section
\ref{sec:conclusions}. 

\section{Case of uniaxial anisotropy \label{sec:uni}}

In the case of surface-supported chains the easy axis of
magnetization, driven by spin-orbit coupling, is either perpendicular
to the chain or along the chain axis, as can be shown by symmetry
arguments \cite{Lazarovits04,footnote2}. For instance, in the case of
Co chains on Pt \cite{Gambardella02} or Co islands and nanoclusters on
Pt \cite{Gambardella03,Rusponi03} the easy axis is perpendicular to
the surface (except in the case of chains at step edges, where there
is a canting induced by the step edge environment)
\cite{Lazarovits04,Shick04,Baud06a,Baud06b,Vindigni06}. The anisotropy barrier is
typically between 1 and 10~meV, strongly dependent on the cluster size
or shape. In this section we assume that the anisotropy tensor
possesses rotational symmetry in the plane perpendicular to the easy axis; this particular {\it Ansatz}
is fulfilled in free-standing chains but is not always valid when the
chain is adsorbed on a surface.  Thus, its approximate validity is
strongly system-dependent and a more general case is investigated in
Section~\ref{sec:multi}. As a rule of thumb, it should be more valid
for chains that are contained in the first surface layer or for chains
of biatomic cross section, compared to chains that are on top of the
surface layer, because in the former case the in-plane asymmetry of
the electron potential is smaller. Other effects that we neglect here
are the possibility of temperature-dependent orientation of the easy
axis that has been observed in certain systems \cite{Kukunin07} and a
possible canting of the anisotropy axis for the edge atoms, which has
been found experimentally and theoretically
\cite{Gambardella02,Lazarovits04,Shick04,Baud06a,Baud06b}. However, at
the end of the section we discuss some effects that could occur due to
increased edge anisotropy.

We proceed with a presentation of our results in the case of
uniaxial anisotropy. We simulated the dynamics of monoatomic chains of
various lengths up to a maximum of 100 atoms, with $K/J=0.01$ and
$0.1$; we chose a damping of $\lambda=0.1$. Values of $K/J>0.1$ are
rather extreme for supported transition-metal chains, but can be
realized for example in cases of molecular magnets (where the exchange
is comparably week), or for unsupported chains of 4d transition metal
atoms \cite{Mokrousov06}. The running time of the simulations varied,
depending on the system, so that statistically reliable results could
be obtained for the lifetime $\tau$.

\begin{figure}
  \includegraphics[width=8cm]{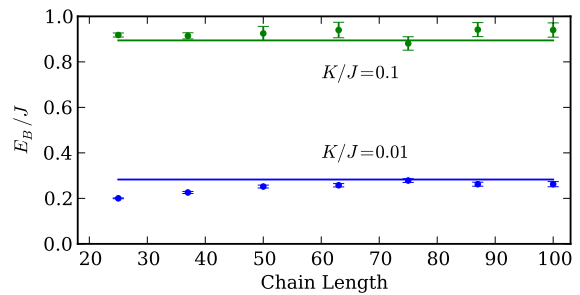}
  \caption{(colour online) Simulation fits to the activation barrier
    height as a function of chain length taking into account a
    uniaxial anisotropy constant $K$. Cases $K/J=0.01$ and $K/J=0.1$
    are shown. The horizontal lines correspond to the analytical
    solution of the continuum approximation
    $\EB/J=2\sqrt{2}\sqrt{K/J}$ and agree rather well with the
    simulation results. For $K/J=0.01$ we see a slight decrease of
    \EB\ at small lengths, indicating the onset of a transition to a
    Stoner-Wohlfarth behaviour as the length becomes comparable to the
    domain wall width. \label{fig:barrier}}
\end{figure}

By fitting the Arrhenius law (\ref{eq:Arrhenius}) to our simulation
results we extracted $\EB$ and verified the hypothesis that the energy
barrier results basically from the formation of a domain wall, if the
chain is not too short. The results presented in
Fig.~\ref{fig:barrier} show that $\EB=\EDW$ holds up to the limit
$L\approx\LDW$; for smaller lengths, a deviation is seen (in
particular in the case $K/J=0.01$, where $\LDW=20$).

\begin{figure*}
  \includegraphics*[width=\textwidth]{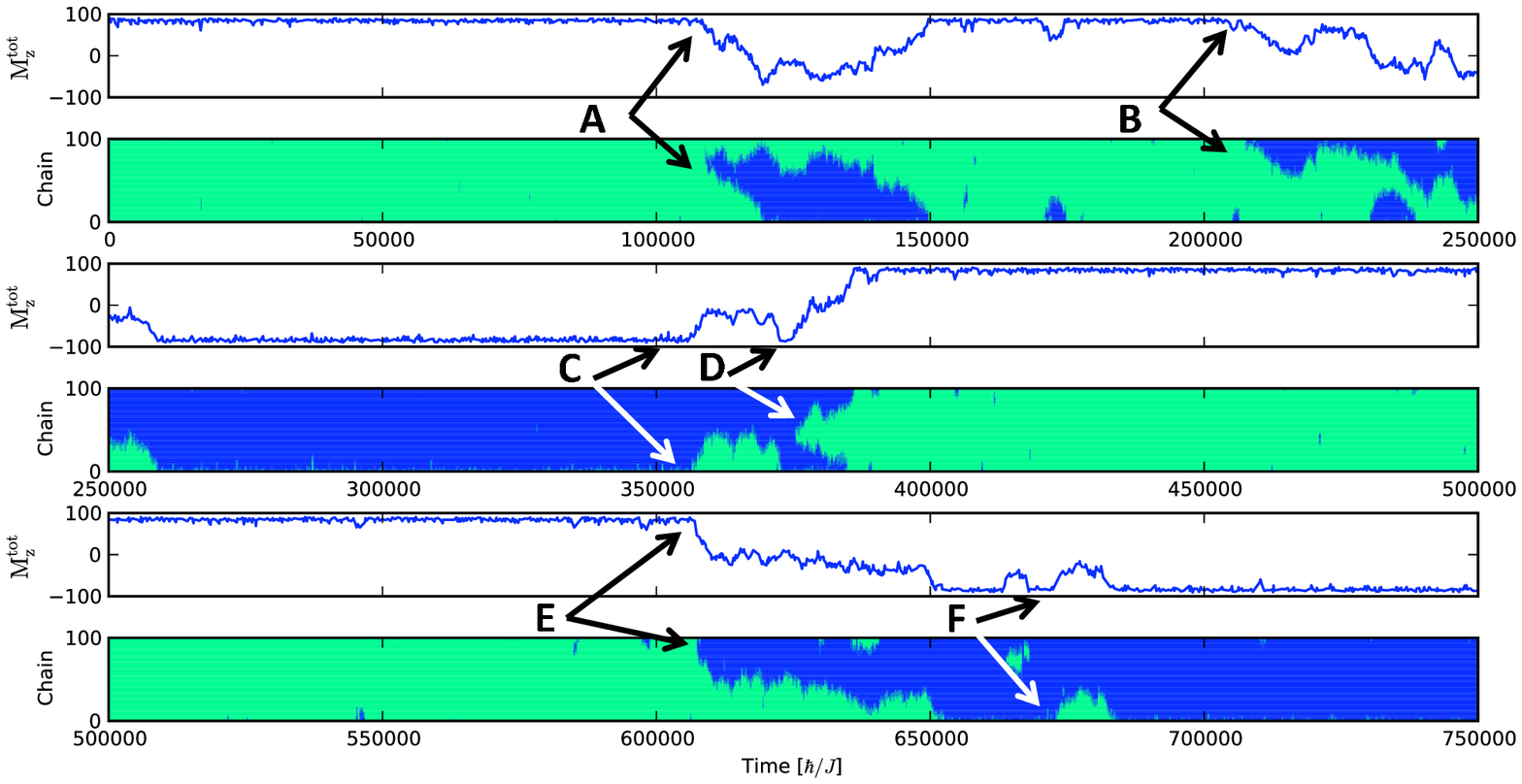}
  \caption{(colour online) Time evolution of simulation snapshots for
    a 100-atom chain with anisotropy $K/J=0.1$. The total simulation
    time interval is broken up in three sub-intervals of duration
    $0.25\times 10^6\ \hbar/J$ each. For each sub-interval, the upper panel
    shows the $z$-component of the total magnetic moment of the chain,
    $M_z^{\text{tot}}(t)$, while the lower, colour-shaded part shows a
    snapshot of the $z$-component of the magnetization at any site in
    time: green for $M_{i;z}(t)>0$, blue for $M_{i;z}(t)<0$. In the
    upper panels the attempts for magnetization reversal are witnessed
    by a significant change of $M_z^{\text{tot}}(t)$, while in the
    lower panels one can see where the attempt is intitiated and how
    the magnetization-reversed region evolves in time. The letters and
    arrows correspond to events that are described in the main part of
    the text (Sec.~\ref{sec:uni}).  \label{fig:timedep}}
\end{figure*}

Next we present in Fig.~\ref{fig:timedep} simulation data on a chain
of 100 spins for $K/J=0.1$ at a temperature of $k_{\rm
  B}T/J=0.11$. The abscissa denotes the simulation time broken up in
three time intervals; for each interval two panels are shown. In the
upper panel, the ordinate shows the chain total magnetic moment along
the easy axis ($z$-axis), $M_z^{\rm tot}$, as a function of
time. Reversal events are evident, indicated as B, D and E, when the
total moment is reversed. Each reversal event is relatively fast,
followed by a longer time interval where the total moment is
approximately stable at $M_z^{\text{tot}}=\pm 85$. However, we also
see at A, C, and F attempted events that failed, with the total moment
returning to its previous state. In the lower panel of
Fig.~\ref{fig:timedep} we see a series of snapshots of the system over
the same time span, visualizing the time-evolution of the
site-dependent magnetization. Here, the ordinate represents the
position of the spin in the chain, $i$, while a colour-code is used
for the sign of the local $M_{i;z}$: green for $M_{i;z}>0$, blue for
$M_{i;z}<0$. We can see how reversal attempts start by nucleation of a
spin-fliped region that either propagates throughout the chain to end
in reversal, as in B, D, and E, or is annihilated in an unsuccessful
attempt as in A, C, and F. We also see that in some attempts the
nucleation starts at the chain edge, as in B, C, E and F, while for
others it starts in the chain interior, as in A and D. Although here
we show a time-interval containing both kinds of nucleation, at not
too high temperatures nucleation is more likely to occur at the edges
than in the interior, because in the latter case two domain walls must
appear simultaneously requiring the system to overcome a twice as high
energy barrier (the latter process is known as soliton-antisoliton
creation and has been studied e.g.\ by Braun \cite{Braun06}). For the
same reason it is statistically rare for more than one nucleation
regions to be present simultaneously, at least as long as $L\ll\xi$,
therefore we consider in our qualitative analysis only single
nucleation events at the edges. 
Our arguments are supported by calculations making use of a transfer-matrix
approach to calculate thermodynamic properties of finite 
anisotropic Heisenberg~\cite{Vindigni06} 
showing that the edge-atoms of chains are subject to significantly stronger 
fluctuations than the atoms in the interior.

The reversal time itself, i.e. the
time that it takes for the domain wall to propagate through the system
once it is created, is in general much shorter than the magnetic-state
lifetime $\tau$. This effect is also seen experimentally by
spin-polarized scanning tunneling microscopy of ferromagnetic
nano-islands \cite{Bode04}.  There, fast reversal occurs during the
scanning process, resulting in an image of seemingly two domains,
while in reality the island is in a mono-domain state.

\begin{figure}
  \includegraphics[width=8cm]{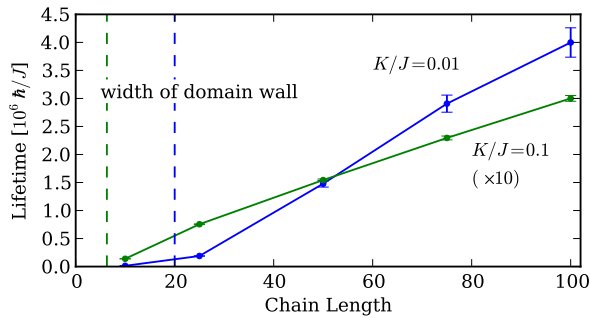}
  \caption{(colour online) Magnetization lifetime $\tau$ as a function
    of chain length $L$ for the cases $K/J=0.01$ and $K/J=0.1$ (the
    latter was scaled up by a factor 10 for better legibility). The linear
    increase of $\tau$ with $L$ is due to the random-walk-type
    propagation of the domain wall after nucleation. The offset of the
    linear behaviour is due to the finite domain wall width. Vertical
    lines indicate the domain wall widths in the two cases ($\LDW=6$
    for $K/J=0.1$, $\LDW=20$ for $K/J=0.01$). \label{fig:prefactor}}
\end{figure}

The domain wall that occurs during nucleation performs a random walk
back-and-forth in the chain, and eventually either returns to its
starting point, resulting in an unsuccesful attempt, or reaches the
other side, resulting in reversal.  The probability for the latter to
happen in a random-walk model is \cite{Cordery81}
$p(L)=1/L$. Therefore, at constant temperature, the prefactor $\tau_0$
of Eq.~(\ref{eq:Arrhenius}) rises linearly with $L$: $\tau_0(L)=\alpha
L$, at least as long as $\LDW\ll L$; this has been found also in
simulations of Ising-model dynamics \cite{Vindigni05}. More accurately,
going beyond the Ising model, we must account for the fact that a
domain wall centered at position $x$ is attracted to the chain edge if
it is too close, i.e., if $x<\frac{1}{2}\LDW$ or
$L-x<\frac{1}{2}\LDW$. Thus the domain wall propagation corresponds to
a random walk only between $\frac{1}{2}\LDW$ and $L-\frac{1}{2}\LDW$,
so that we should observe an offset in the linear behaviour:
\begin{equation}
\tau_0(L)\propto (L-\LDW),\ \mbox{for}\ L>\LDW.  
\label{eq:randomwalk}
\end{equation}
This rule is reproduced by the results of our simulations, as shown in
Fig.~\ref{fig:prefactor} where the calculated lifetime is plotted against the chain length.
The offset in the 1st curve ($K/J=0.01$) can
be seen at lengths smaller or comparable to the domain wall width (up
to 20-25 atoms). Then, for $L\approx\LDW$, the switching mechanism is
intermediate between nucleation and Stoner-Wohlfarth-like coherent
rotation, and a kink in the curve $\tau(L)$ is seen. The offset is
much reduced in the 2nd curve of Fig.~\ref{fig:prefactor} where
$K/J=0.1$, because the increased anisotropy leads to smaller
$\LDW\approx 6$. A deviation from the rule (\ref{eq:randomwalk}), in
particular a saturation, is expected to begin as the chain length
becomes too large approaching the correlation length. Then
simultaneous occurence of more that one nucleation becomes
statistically more and more relevant, until eventually the
space-integrated magnetization vanishes at all times.  Note that this
linear increase (\ref{eq:randomwalk}) of the lifetime with system size
is in marked difference to the behaviour of a Stoner-Wohlfarth-type
three-dimensional particle, where the system size (number of atoms
$n$) enters multiplicatively in the Arrhenus barrier, $\EB=nK$,
causing an exponential dependence of $\tau$ on $n$. 
On the other hand, in the regime of validity of Eq.~(\ref{eq:randomwalk}) ($L>\LDW$), the Arrhenius law (\ref{eq:Arrhenius}) is still valid, but with $\EB$ a constant (i.e., independent of the number of atoms).

We close this section by an observation concerning the anisotropy of
the edge atoms of the chain. It is well-known that, due to the
different environment (only one neighbour) the anisotropy tensor of
edge atoms, $\mat{K}_{\text{edge}}$, can differ from the one in the
chain interior, in both the direction of the principal axes and in the
anisotropy strength, $K_{\text{edge}}$. These effects can affect the
nucleation barrier as well as the nucleation frequency. However, there
can be in principle an additional effect. Under ``normal'' conditions
a domain wall close to the edge is further attracted toward the edge
and in the end expelled form the chain and annihilated, but if
$K_{\text{edge}}$ becomes large enough, it will act as a barrier
against the annihilation. Then the domain wall will be trapped in the
interior of the chain for a longer time, eventually overcoming the
barrier due to thermal fluctuations and becoming
annihilated. Naturally, in such a situation the nucleation is also
more difficult to achieve. We derived analytically an approximate
condition for such an edge barrier by considering the energy as a
function of rotation of the edge spin and treating the rest of the
chain in the continuum limit (see Appendix B). The condition reads:
\begin{equation}
 K_{\text{edge}} > K + \EDW/4
\label{eq:edge}
\end{equation}
and is approximately reproduced by simulations that we performed.

\section{Case of triaxial anisotropy  \label{sec:multi}}

As we commented earlier, the magnetic anisotropy of a chain on a
surface is never really uniaxial, due to the asymmetry of the
environment. We take this into account in the simplest way by
introducing three eigenvalues of the anisotropy tensor, 
\begin{equation}
\mat{K}= \left(
\begin{array}{ccc}
K_{x} & 0 & 0 \\
0 & K_{y} & 0 \\
0 & 0 & K_{z} \\
\end{array}
\right),
\end{equation}
corresponding so to say to an easy, medium, and hard axis, so that the
associated energy expression becomes
$-\vc{M}\cdot\mat{K}\vc{M}=-K_xM_x^2 - K_yM_y^2 - K_z M_z^2$; the site
index $i$ has been suppressed here, and it is assumed that all sites
have the same anisotropy. Without loss of generality we fix $K_z\geq
K_y \geq K_x = 0$ (the latter equality can be adjusted by a shift of
the energy zero that does not affect the equations of motion, while
the order of the inequalities will not affect the statistical
results on the lifetime, merely the path that the spins follow to
flip). Thus, in effect the two energy minima of the single-ion
anisotropy are along the $\pm z$-axis, separated by a low anisotropy
barrier $\low=(K_z-K_y)$ if the spin rotates in the $y$-$z$-plane or
by a high anisotropy barrier $\high=K_z$ if the spin rotates in the
$x$-$z$-plane; effects of the exchange $J$ come on top of this as
before.

The question that we seek to answer is how the presence of the second
anisotropy barrier affects the magnetization lifetime. To make the
question more concise, we wish to compare the lifetime in the
previously calculated uniaxial case, where $\high=\low=K$, to the case
$\high>\low=K$. The minimal barrier for domain wall formation in long
chains or for coherent rotation in short chains remains
$2\sqrt{2JK}$ or $LK$ respectively, so that the exponent $\EB$ in the
Arrhenius law (\ref{eq:Arrhenius}) should not be affected. However,
the attempt frequency $\tau_0^{-1}$ is expected to change, as the
energy lanscape of the paths that connect the two minima is now
different.  For example, in the case of uniaxial anisotropy both
Bloch-type and N\'{e}el-type walls are equal in energy, while in the
tri-axial case one of the two types will correspond to a higher
energy; this must give a difference in entropy.

\begin{figure}
  \includegraphics[width=8cm]{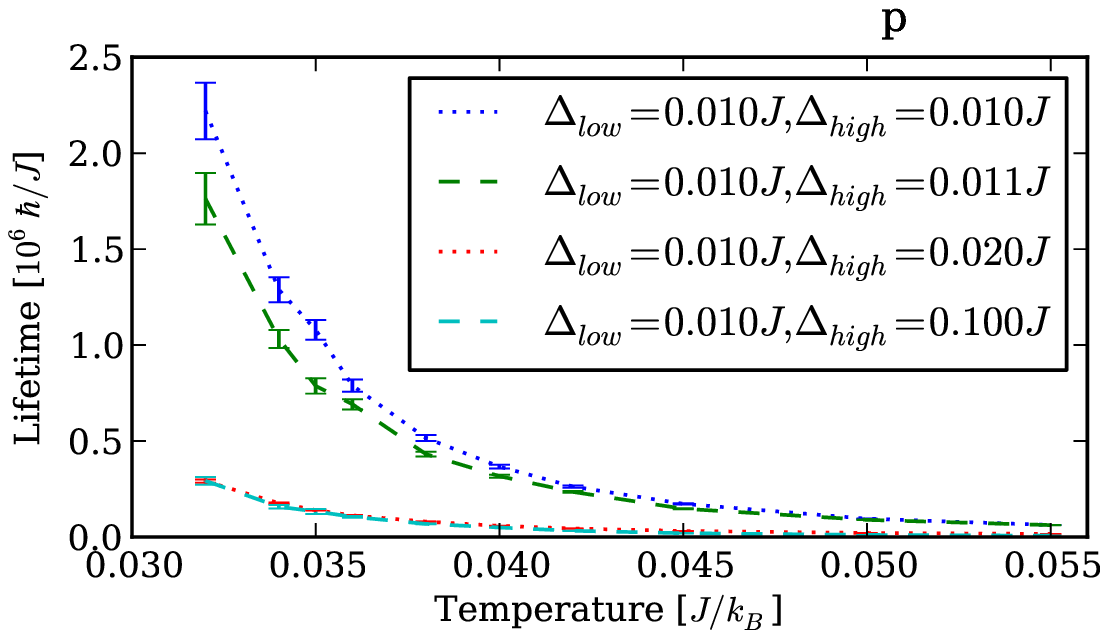}
  \includegraphics[width=8cm]{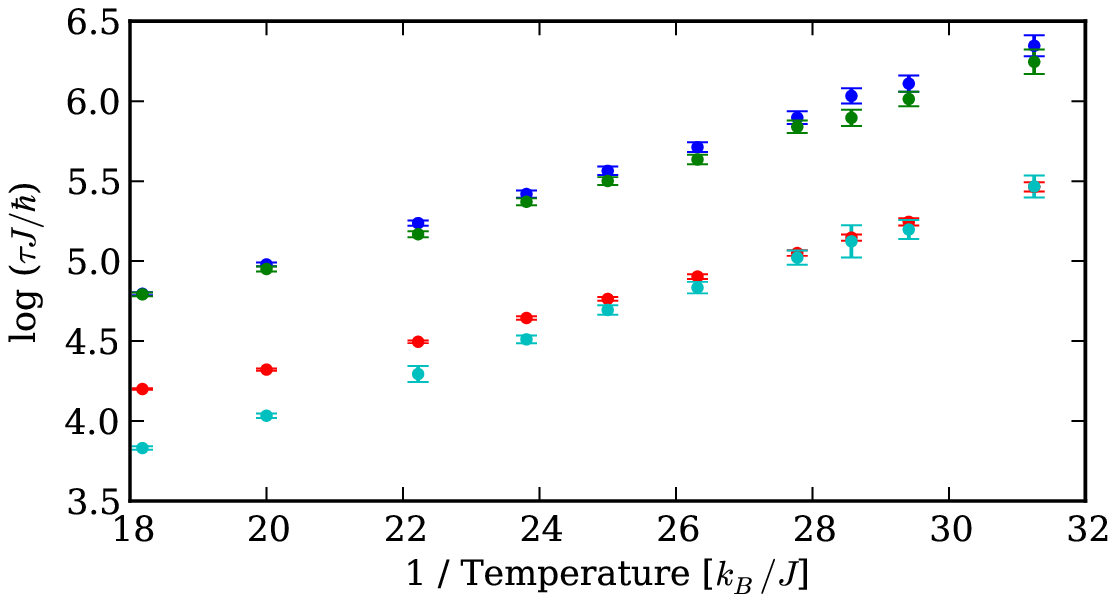}
  \caption{(colour online) Magnetization lifetime as a function of
    temperature for a 100-atom chain for several single-ion anisotropy
    barriers, including triaxial anisotropy (top, linear scale;
    bottom: logarithm of the lifetime {\it vs.} inverse
    temperature). The low barrier, $\low=0.01 J$, is kept constant,
    while \high\ is varied between \low\ and $10 \low$. A significant
    reduction of the lifetime is observed even at a moderate
    difference of 10\% between \high\ and \low. At very high difference
    a saturation is seen. The activation barrier is in all cases the
    same, equal to the domain wall energy, $\EB=2\sqrt{2J\low}$, as
    can be seen in the lower panel where all data sets show the same
    slope.\label{fig:twobarrier}}
\end{figure}

As it turns out, for a given \low, the lifetime \emph{decreases} with
increasing $\high$; our findings are summarized in
Fig.~\ref{fig:twobarrier}. This is at a first glance
counter-intuitive: the fluctuating force enforces a random walk,
regulated by the barriers, that has to climb up from an energy minimum
to a maximum or saddle point before descending to the opposite
minimum. Increasing $\high$ means that part of the lanscape is visited
with smaller frequency, so that there are less escape paths. Then part
of the random-walk steps would be, so to say, waisted in attempts to
climb up the high barrier. However, contrary to the case of a confined
particle, the latter argument is not valid in the case of spins,
because of the precessional motion. The spin trajectory has a
component following the direction of the effective field (toward the
energy minimum), given by the damping term of Eq.~(\ref{eq:LL}), but
the stronger, precessional part drives the spin perpendicular to the
energy gradient. Therefore, whenever the fluctuating field forces the
spin in the direction of the high barrier, the precession drives it in
the direction of the low barrier. In this way, the random-walk steps
in the direction of the high barrier are not waisted but converted into
random-walk steps in the direction of the low barrier. The shape of
the resulting trajectory approximates an ellipsis with its long axis
along the low-barrier direction. The random walk becomes in a sense
more and more one-dimensional, as the difference $\high - \low$
increases, and it is known that random walks in one dimension
propagate faster away from the starting point than in two
dimensions. Therefore the magnetization lifetime becomes shorter. In
the limit $\high \gg \low$, the situation resembles that of an
$y$-$z$-easy-plane model with a hard axis in the $x$ direction, which
obeys the dynamics of precessional switching, long known in
micromagnetics \cite{Serpico03}. At this limit, even a small deviation
of the spin toward the $x$ direction due to the fluctuating force will
cause a lengthy precession around $x$, in the $y$-$z$ plane, which can
even last a few rotations if the damping is very low.

\begin{figure}
  \includegraphics*[width=8.5cm]{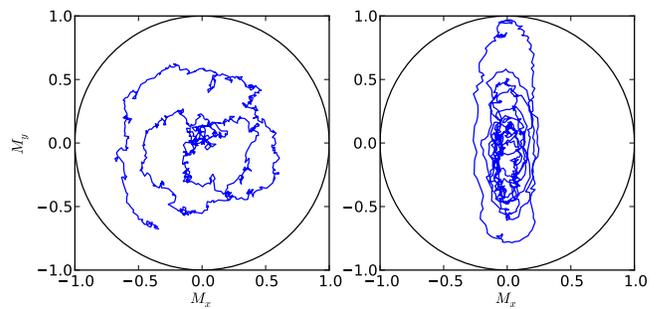}
  \caption{(colour online) Trajectory of a single spin under
    anisotropy and fluctuating force. Left: Uniaxial anisotropy,
    $K_x=K_y=0$, $K_z=K>0$ ($\high=\low=K$). Right: $K_x=0$, $K_y=9K$,
    $K_z=10K$ ($\high=10K,\low=K$). Both simulations correspond to the
    same total simulation time starting from the state
    $M_x=M_y=0$. Evidently, in the presence of different barriers in
    $x$ and $y$ directions (right panel) the spin reaches faster the
    turning point at the edge of the circle ($M_x^2+M_y^2=1$, $M_z=0$).
    \label{fig:trajec}}
\end{figure}
The behaviour is visualized for a single spin in
Fig.~\ref{fig:trajec}. On the left panel we see a top-view of the
trajectory of a spin under a uniaxial anisotropy, with $K_x=K_y=0$,
$K_z=K>0$, starting from the state $M_x=M_y=0$, $M_z=1$. The spin is
driven away from equilibrium by a fluctuating field, and the
trajectory is, on the average, isotropic in $x$ and $y$. On the right
panel we see the trajectory for the same simulation time and under the
same conditions, with the exception that here $\low=K$,
$\high=10K$. The point is not just that the trajectory is elliptical
instead of isotropic, but also that in the same time interval the spin
has managed to reach closer to a ``turning point'' of the barrier,
where $M_x^2+M_y^2=1$, $M_z=0$.

In the case of chains (Fig.~\ref{fig:twobarrier}) we observe in the
top panel a drop of the lifetime with increasing $\high$, while
keeping $\low=0.01J$ fixed. The curves show the lifetime as a function
of temperature. At all temperatures, the longest lifetime is observed
for $\high = \low$, and it is evident that even a small increase of
10\% in $\high$ can affect the lifetime quite noticeably. When \high\
is twice the value of \low, the lifetime is reduced by an order of
magnitude. However there appears also to be a saturation: further
increase of $\high$ to nine times the value of \low\ makes no
significant difference any more for the lifetime.

In the lower panel, where the same simulation data are shown in a
$\log \tau${\it vs.}~$1/T$ plot, the Arrhenius behaviour as well as
the dependence on $\high-\low$ becomes easier to see also at higher
temperatures; evidently the activation barrier is the same in all
cases (and equal to $2\sqrt{2J\low}$), as can be seen from the fact
that the slope of $\log \tau${\it vs.}~$1/T$ does not change. Note
that at low temperatures the data have larger error bars due to
the fewer switches that occur in a given simulation time interval.

\section{Realistic parameters and results for C\lowercase{o} chains on
  P\lowercase{t} \label{sec:real}}

Now we relate our general analysis to realistic systems. For the
typical system of supported monoatomic Co chains and nanoislands, for
example, values of $J$ of the order of 20-60~meV on Cu, Au and Pt
surfaces have been calculated and reported in the literature
\cite{Mavropoulos10,Sipr06,Minar06,Frota06}, depending on structural
parameters such as island or chain geometry, size and
substrate. Similar is the case with Fe nanostructures. The anisotropy,
on the other hand, is found in experiments to vary by two orders of
magnitude depending on structural parameters \cite{Gambardella03},
from the sub-meV range for large clusters up to values as high as
10~meV for single adatoms; for monoatomic Co chains on Pt(997) step
edges, a value of 2~meV has been found by experiments
\cite{Gambardella02}, while the value drops to 0.34~meV for biatomic
chains. Clearly, the anisotropy is much more sensitive than the
exchange coupling. Concerning the appearance of two anisotropy
barriers, \low\ and \high, calculations \cite{Conte08} on Co
monoatomic chains on Pt(111) yield $\low=0.8$~meV and $\high=1$~meV;
this 25\% difference between \low\ and \high\ can play a significant
role on the lifetime, as our calculations of Sec.~\ref{sec:multi}
show.

Following the above observations we are in a position to make an
order-of-magnitude prediction for the lifetime of Co monoatomic chains
on Pt. We assume values of $J=50$~meV, $\low/J=0.016$
$\high/J=0.02$. The only parameter that is at this point arbitrary is
the damping, here chosen to be $\lambda=0.1$, which is not actually
known for the case of Co monoatomic chains on Pt. In permalloy, for
example, the damping is one order of magnitude smaller
\cite{Nahrwold10}. However, Pt is characterized by a much stronger
spin-orbit coupling, which can enhance the damping (for example, FePt
nanoparticles are reported to have a damping parameter as high as 0.76
\cite{Kotzler07}). Our current choice has to be corrected once there
is more experimental or theoretical data, 
including exchange parameters beyond first neighbours and 
an account of the anisotropic exchange (Dzyaloshinskii-Moriya interaction),
on the particular type of
systems. Keeping this in mind, we find for a 50-atom chain an inverse
attempt rate of $\tau_0\approx 0.6$~ps and a barrier of
$\EB=2\sqrt{2J\low}\approx 18$~meV to be substituted in the Arrhenius
law (\ref{eq:Arrhenius}). Assuming detection of the reversal by a
scanning tunneling microscope (see e.g. Refs. \cite{Bode04} and
\cite{Krause07}), and given that this method has a relatively low time
resolution of the order of msec \cite{Loth10}, we consider a lifetime
of 1~second as reasonable for a reliable experimental analysis of
spontaneous reversal. Then our simulation suggests that experiments on
50-atom-long chains require temperatures of approximately 10~Kelvin or
lower.

\section{Conclusions \label{sec:conclusions}}

We have performed classical simulations of atomistic spin dynamics for
supported monoatomic magnetic chains, with focus on the magnetization
lifetime between temperature-induced subsequent magnetization
reversals at chain lengths below the correlation length. The
calculations were carried out by integrating the Landau-Lifshitz
equation of motion of classical spin systems in the presence of
stochastic forces.

In summary, the main findings of this work are the following. (i) For
long chains (longer than the width of a domain wall), the lifetime is
governed by an Arrhenius law (\ref{eq:Arrhenius}), with a barrier
equal to the domain wall energy. For short chains, the barrier
eventually becomes equal to the anisotropy energy. (ii) For long
chains, the mechanism for reversal is governed by nucleation of a
region of reversed magnetization at the chain boundary, forming a
domain wall, and propagation of the domain wall through the chain via
a random-walk-like procedure. This yields a linear dependence of the
lifetime on the chain length for chains longer than the domain-wall
width, contrary to an exponential dependence that would be expected
for a Stoner-Wohlfarth type of system. (iii) In case of a triaxial
anisotropy tensor with non-degenerate eigenvalues, i.e., with a low
and a high anisotropy barrier for magnetization reversal, the lifetime
is reduced compared to the case of only a low barrier. This effect is
related to the precessional motion of the spin, which tends to
transform fluctuations in the high-barrier direction into fluctuations
in the low-barrier direction. A triaxial anisotropy should always be
present in surface-supported chains, as can be argued on the grounds
of symmetry of the system.

As far as fundamental science is concerned, the most important
question to be answered by this type of calculations is perhaps the
one of the limitations in size, time and temperature, that the
classical spin dynamics is still a valid approximation, especially in
metallic systems. In principle, all ingredients that enter the
Landau-Lifshitz equations are experimentally accessible; even more so
for chains, where the geometry is simple enough so as not to introduce
further arbitrariness. The result of the simulations, in particular
the magnetization lifetime, is also accessible by experiment. A
systematic study in this direction will be highly interesting.

\section*{Acknowledgements}

We are greatful to Dr. Riccardo Hertel for discussions on the physics
of magnetization dynamics, Dr. Laszlo Szunyogh for discussions on the
form of the anisotropy tensor in surface-supported chains, and
Prof. Christian Schr\"oder for discussions on the methodology of
atomistic spin dynamics. This work has been supported in part by FP7
EU-ITN FANTOMAS. S. L. wishes to thank the Alexander von Humboldt
Foundation for a Feodor Lynen Fellowship and also Prof. D. L. Mills
for hospitality at the UC-Irvine.

\appendix

\section{Some details on the method of calculation}
Equation (\ref{eq:LL}) is a stochastic differential equation with
multiplicative noise which is interpreted in the Stratonovich sense.
The fluctuating force $\vc{f}_i(t)$ has white-noise properties,
i.e.\ (i) its time-average vanishes: $\langle\vc{f}_i\rangle=0$; (ii)
it is fluctuating much faster than the characteristic precession time
so that time-separated and space-separated fluctuations are
decorrelated:
$\langle[\vc{f}_i(t)]_{\alpha}[\vc{f}_j(t')]_{\beta}\rangle
=\delta_{ij}\delta_{\alpha\beta}\delta(t-t')$, where
$\alpha,\beta=x,y,z$ denote the vector components and $i,j$ are atomic
sites; and (iii) its amplitude $\epsilon$ is connected to the damping
via the fluctuation-dissipation theorem: $\epsilon^2=2\lambda k_{\rm B} T$
with $k_{\rm B}$ the Boltzmann constant. Eq.~(\ref{eq:LL}) is integrated via a
weak Runge-Kutta method which was suggested by Milstein and Tretyakov
\cite{Milstein97}.  The error scales with the step size as
$\mathcal{O}(h^4+ \epsilon^2 h^2)$. Further information can be found
in Ref.~\cite{Antropov97}.

In practice, at each Runge-Kutta step and for each spin at site $i$, a
fluctuating field $\vc{f}_i$ is calculated by a set of random numbers
obeying an appropriate distribution so that the above requirements
(i-iii) are fulfilled. Requirement (iii), in particular, guarantees
that the fluctuating field intensity corresponds to the particular
simulation temperature $T$. The field $\vc{f}_i$ acts in addition to
the effective field $\Beff_i$ and to the damping term as shown in
Eq.~(\ref{eq:LL}). The random nature of $\vc{f}_i$ comes into notice in the noise
observed in the time-dependent magnetiation in Fig.~\ref{fig:timedep}.
For the calculation of $\tau(T)$, during a sufficiently long
simulation time at a fixed temperature, successive time intervals
$\Delta t_i$ between $N$ successive reversal events (as the ones seen
in Fig.~\ref{fig:timedep}) are recorded, and at the end averaged as
$\tau(T)=\sum_{i=1}^N\Delta t_i/N$.

\section{Justification of Expression~(\ref{eq:edge})}

Consider a domain wall where the magnetic moment of each atom $i$ has
an angle $\theta_i$ with respect to the easy axis (the easy axis is
taken to be the same for all atoms). Suppose that the wall is not at
the interior of the chain, but close to the edge, and that it is
formed by constraining the moment of the first atom (labelled ``0'')
by $\theta_0$; we accept the boundary condition at the deep interior
of the chain that $\theta_i=0$ for $i\rightarrow\infty$. Then, a
boundary condition at the edge $\theta_0=\pi$ corresponds to a fully
formed domain wall, while $\theta_0=0$ corresponds to the
ferromagnetic ground state.

Now we take a fixed $\theta_0$, $0<\theta_0<\pi$. The resulting
structure will have the form of a domain wall that has been abruptly
cut at $i=0$. 
If we assume in the discrete model a nearest neighbour distance of length $a$ 
and employ the continuum approximation \cite{Aharoni00}, then the site index $i$ changes to a continuous 
variable $x/a$ and we obtain an  explicit energy functional
for a chain starting with a restricted angle of $\theta_0$ at $x=0$:
\begin{equation}
E(\theta_0)=\int_{0}^\infty dx \, 
\left(\frac{1}{2} Ja \,\left(\frac{d\theta}{dx}\right)^2 + \frac{K}{a}\, \sin^2(\theta(x))\right) .
\end{equation}
Minimizing the energy functional under the boundary conditions
$\theta(0)=\theta_0$ and $\theta(\infty)=0$ leads to an energy of the
domain wall of
\begin{equation}
\EDW(\theta_0)=\sqrt{2JK} (1-\cos\theta_0), \label{eq_EDW}
\end{equation}
which will be used as an approximation to the discrete model.

Next we increase the anisotropy of only the edge atom from $K$ to
$K_{\text{edge}}$. This change affects only the energy of the edge
atom, changing it by $-K_{\text{edge}}\cos^2\theta_0+K\cos^2\theta_0$,
while the configuration remains the same. We have now an energy 
\begin{equation}
E(\theta_0) = \sqrt{2JK}(1-\cos\theta_0) - K_{\text{edge}}\cos^2\theta_0+K\cos^2\theta_0.
\label{eq:B3}
\end{equation}
For the domain wall to be trapped in the interior of the chain, this
expression must have a maximum for $0<\theta_0<\pi$, corresponding to
the high-point of an energy barrier upon rotation of the edge
spin. Differentiation of (\ref{eq:B3}) with respect to $\theta_0$ shows that such a
maximum exists under the condition
\begin{equation}
K_{\text{edge}}-K > \sqrt{JK/2}, 
\end{equation}
which is equivalent to Expression~(\ref{eq:edge}).


\begin{thebibliography}{99}
\bibitem{footnote} Contrary to chains, calculations show that in
  nanoislands and nanoclusters the inter-atomic exchange interactions
  fluctuate strongly as a function of cluster shape and size and of
  the position of the atoms in the cluster \cite{Mavropoulos10}.
\bibitem{footnote2} This is strictly true only if we neglect a
  possible weak variation of the single-ion anisotropy due to the
  geometric arangement of the nearest neighbours on the
  substrate. E.g. the triangular geometry of (111) surfaces can cause
  a further tilting of the anisotropy axis.
\end{thebibliography}
\end{document}